\documentclass[12pt]{article}
\newcommand{\sect}[1]{\setcounter{equation}{0}\section{#1}}
\newcommand{\eq}{\begin{equation}}
\newcommand{\eqa}{\begin{eqnarray}}
\newcommand{\en}{\end{equation}}
\newcommand{\ena}{\end{eqnarray}}
\newcommand{\enn}{\nonumber \end{equation}}

\def\sk{\vskip .4cm}
\def\noi{\noindent}
\def\om{\omega}
\def\al{\alpha}

\def\ga{\gamma}
\def\Ga{\Gamma}

\let \part\partial

\def\unquarto{{1 \over 4}}

\def\unmezzo{{1 \over 2}}
\def\epsi{\varepsilon}
\def\we{\wedge}

\def\de{\delta}

\def\part{\partial}

\def\sk{\vskip .4cm}

\def\noi{\noindent}

\def\X0{X^0}

\def\om{\omega}

\def\al{\alpha}
\def\ga{\gamma}

\def\unquarto{{1 \over 4}}
\def\unmezzo{{1 \over 2}}
\def\epsi{\varepsilon}

\def\we{\wedge}

\def\de{\delta}

\def\square{{\,\lower0.9pt\vbox{\hrule \hbox{\vrule height 0.2 cm
\hskip 0.2 cm \vrule height 0.2 cm}\hrule}\,}}

\def\epsilonbar{{\bar \epsilon}}

\def\Lcal{{\cal L}}

\def\Phi{\phi}
\def\rf{{\rm f}}

\def\fbaralup{\overline{\rf}^\alpha}
\def\fbaraldo{\overline{\rf}_\alpha}

\def\westar{\we_\star}

\def\omtilde{\tilde \om}
\def\Vtilde{\tilde V}
\def\Ttilde{\tilde T}

\def\rtilde{\tilde r}
\def\epsitilde{\tilde \epsi}

\def\psibar{\bar \psi}
\def\chibar{\bar \chi}
\def\rhobar{\bar \rho}
\def\Om{\Omega}

\begin{document}

\begin{titlepage}
\rightline{DISTA-UPO/09}
%\rightline{hep-th/9509031}
\rightline{February 2009} \vskip 2em
\begin{center}{\bf NONCOMMUTATIVE SUPERGRAVITY IN D=3 AND D=4}
\\[3em]
{\bf Paolo Aschieri}${}^{1,2}$ and {\bf Leonardo Castellani}${}^{2}$ \\ [3em] {\sl ${}^{1}$
Centro ``Enrico Fermi", Compendio Viminale, 00184 Roma, Italy \\
[1em] ${}^{2}$ Dipartimento di Scienze e Tecnologie avanzate and
\\ INFN Gruppo collegato di Alessandria,\\Universit\`a del Piemonte Orientale,\\ Via Bellini 25/G 15100
Alessandria, Italy}\\ [1.5em]
\end{center}

\begin{abstract}
We present a noncommutative  $D=3$, $N=1$ supergravity, invariant
under diffeomorphisms, local $U(1,1)$ noncommutative $\star$-gauge
transformations and local $\star$-supersymmetry. Its commutative
limit is the usual $D=3$ pure supergravity, without extra fields.

A noncommutative deformation of $D=4$, $N=1$ supergravity is also
obtained, reducing to the usual simple supergravity in the
commutative limit. Its action is invariant under diffeomorphisms
and local $GL(2,C)$ $\star$-gauge symmetry. The supersymmetry of
the commutative action is broken by noncommutativity. Local
$\star$-supersymmetry invariance can be implemented in a
noncommutative  $D=4$, $N=1$ supergravity with chiral gravitino
and complex vierbein.

\end{abstract}

\vskip 6cm \noi \hrule \vskip.2cm \noi {\small
leonardo.castellani@mfn.unipmn.it\\ aschieri@to.infn.it}

\end{titlepage}

\newpage
\setcounter{page}{1}

\sect{Introduction}

Gravity theories on $D=4$ twisted spaces have been constructed in
the past in the context of particular quantum groups
\cite{Castellaniqgrav} and more recently in the twisted
noncommutative geometry setting \cite{Chamseddine,Wessgroup,
Estradaetal}. In this setting the deformed theory is invariant
under $\star$-diffeomorphisms, but in \cite{Wessgroup} no gauge
invariance on the tangent space (generalizing local Lorentz
symmetry) is incorporated, and therefore coupling to fermions
could not be implemented. A local symmetry, enlarging the local
$SO(3,1)$ symmetry of $D=4$ Einstein gravity to $GL(2,C)$, has
been considered in  the approach of Chamseddine
\cite{Chamseddine}. The resulting theory has a complicated
classical limit, with two vielbeins (or, equivalently, a complex
vielbein). Noncommutative gravities in lower dimensions have been
studied in \cite{MilanogroupG2} (D=2) and in
\cite{Banados,MilanogroupG3} (D=3).

In \cite{AC1} we have proposed a noncommutative
gravity, coupled to fermions, and reducing in the commutative limit to
ordinary gravity + fermions, without extra fields (in particular
without an extra graviton).
This is achieved by imposing
a noncommutative charge conjugation
 condition on the bosonic fields, consistent with
the $\star$-gauge transformations.
One can also impose a noncommutative generalization of the Majorana condition on
the fermions, compatible with the $\star$-gauge transformations.

In this paper we present the noncommutative extensions of locally
supersymmetric $D=3$ and $D=4$ gravity theories. The
noncommutativity is given by a $\star$-product associated to a
very general class of twists. This $\star$-product can also be
$x$-dependent. The deformed supergravity actions are constructed
with a cyclic integral.  As a particular case we obtain
noncommutative supergravities where noncommutativity is realized
with the Moyal-Groenewald $\star$-product.

For $D=3$ the situation is easier, since in three dimensions
gravity becomes essentially a Chern-Simons gauge theory.
The noncommutative extension of a particular $AdS(3)$ supergravity
in three dimensions has been studied in \cite{MilanogroupSG3}.

Here we discuss $D=3$, $N=1$ supergravity without cosmological
term. The noncommutative geometric action is constructed directly
by generalizing the usual $D=3$ supergravity action, without
reference to the Chern-Simons action. The noncommutative theory is
invariant under diffeomorphisms, local $U(1,1)$ $\star$-gauge
symmetry and $\star$-supersymmetry.

We then propose an action for a noncommutative deformation of
 $D=4$, $N=1$ supergravity, invariant under diffeomorphisms and local $GL(2,C)$
$\star$-gauge transformations, but without $\star$-supersymmetry.
In this case noncommutativity breaks the local supersymmetry of
the commutative theory. The commutative $\theta \rightarrow 0$
limit is the usual $D=4$, $N=1$ simple supergravity, with a
Majorana gravitino.

We can obtain  local $\star$-supersymmetry invariance of the
noncommutative action if we impose a Weyl condition on the
fermions, rather than a Majorana condition. This leads to a
noncommutative supergravity whose $\theta \rightarrow 0$ limit is
a chiral $D=4$, $N=1$ supergravity with two vierbein fields (or a
complex vierbein) and a left-handed gravitino.

The paper is organized as follows. In Section 2 we discuss
three dimensional noncommutative simple supergravity,
in first order formalism.
In Section 3 we present the index-free formulation of
usual $D=4$, $N=1$ supergravity, exploiting the Clifford algebra representation
of boson fields, thus preparing the ground for its
noncommutative extension. In this setting the supersymmetry of
the action becomes quite easy to prove. In Section 4 we consider noncommutative
 first order $D=4$, $N=1$ supergravity, and prove its local
 $\star$-invariances.  Section 5 contains some conclusions.
 In Appendix A we collect a few useful results of twist differential geometry.
 Conventions, $D=3$ and $D=4$ gamma matrices properties are summarized in Appendices B and C.

\sect{Noncommutative $D=3$, $N=1$ supergravity }

\subsection{Action}

Using the $*$-exterior product of twist differential geometry (see
Appendix A), we extend the usual action of $D=3$, $N=1$ supergravity
to its noncommutative version. In index-free notation:

\eq S =  - 2  \int Tr [R (\Om) \westar V+ i \rho \westar \psibar ]
\label{actionD3}
\en

\noi The fundamental fields are the 1-forms $\Om$  (spin
connection), $V$ (vielbein) and gravitino  $\psi$. The curvature 2-form $R$ and the
gravitino curvature $\rho$ are defined by
 \eq
  R = d\Om - \Om \westar
\Om, ~~~~~ \rho \equiv D\psi = d\psi - \Om \westar \psi
\en
\noi with
 \eq \Om =  \unquarto \om^{ab} \ga_{ab} + i \om 1, ~~~~~V = V^a \ga_a + i v 1~~~~~~
\en
\noi and thus are $2 \times 2$ matrices with spinor indices,
see Appendix B for $D=3$ gamma matrix conventions and useful relations. The Dirac conjugate is
defined as usual: $\psibar = \psi^\dagger \ga_0$. Then $(D\psi) \westar \psibar$ is also a
matrix in the spinor representation, and the trace  $Tr$ is taken on this representation.
Using the $D=3$ gamma matrix  identity:
\eq
Tr (\ga_a \ga_b \ga_c) = -2 \epsi_{abc}
  \en
 \noi allows to rewrite the action in terms of component fields:
   \eq
   S = \int  R^{ab} \westar V^c \epsi_{abc} + 4 r \westar v  + 2i  \psibar \westar \rho
    \label{actionD3comp}
   \en
\noi with
\eq
 R \equiv \unquarto R^{ab} \ga_{ab}+ i r 1,
\en
 \noi and
 \eqa
  & & R^{ab} = d\om^{ab}  - \unmezzo \om^{a}_{~c} \westar  \om^{cb} + \unmezzo \om^{b}_{~c} \westar  \om^{ca} - i (\om^{ab} \westar \om
  + \om \westar \om^{ab}),\\
  & & r = d\om - i \om \westar \om - {i \over 8} \om^{ab} \westar \om_{ab}
  \ena

 \subsection{Hermiticity conditions and reality of the action}

Hermiticity conditions can be imposed on $V$ and $\Om$:
  \eq
 \ga_0 V \ga_0 = V^\dagger,~~~ \ga_0 \Omega \ga_0 =
 \Omega^\dagger \label{hermcond}
 \en
 \noi Moreover it is easy to verify that:
 \eq
 \ga_0 R \ga_0 = R^\dagger,~~~\ga_0 [\rho \westar \psibar] \ga_0 = [\psi\westar \rhobar]^\dagger
\en
\noi with \eq
 \rhobar = d \psibar - \psibar \westar \Om
 \en
 \noi Note also that up to boundary terms
 \eq
    \int Tr [\rho \westar \psibar] =  \int Tr [\psi \westar \rhobar] = -\int \psibar \westar \rho =
    - \int \rhobar \westar \psi
    \en
    \noi where we have used the cyclicity of $Tr$ and the graded cyclicity
    of the integral. For example the first equality holds because
     \eq
      \int Tr [ \rho \westar \psibar ]=
      \int Tr [ d (\psi \westar \psibar) + \psi \westar \rhobar]
      \en
These formulae can be used to check that the action
(\ref{actionD3}) is real.

The hermiticity conditions (\ref{hermcond}) imply that the
component fields $V^a$, $v$, $\om^{ab}$, $\om$ are real.

 \subsection{Field equations}

  \noi  Using the cyclicity of $Tr$ and the graded cyclicity of the integral  in (\ref{actionD3}),
 the variation of $V$ , $\Omega$ and $\psibar$ yield respectively the noncommutative
  Einstein equation,  torsion equation and gravitino equation in index-free form:
 \eqa
  & & R=0 \\
  & & dV- \Om \westar V - V \westar \Om - i\psi \westar \psibar = 0 \label{torsioneqD3}\\
  & & \rho =0
  \ena
   The noncommutative torsion two-form is defined by:
  \eq
  T \equiv T^a \ga_a + i t 1 \equiv  dV -\Om \westar V - V \westar \Om
  \en
  \noi or, in component fields:
  \eqa
   T^a &=& dV^a -\unmezzo (\om^a_{~b} \westar V^b - V^b \westar  \om^a_{~b} )
          + {i \over 4} \epsilon^{abc} (\om_{bc} \westar v + v \westar \om_{bc} ) \nonumber \\
       & &   -i(\om \westar V^a + V^a     \westar \om)\\
        t &= &dv-{i \over 4} \epsilon_{abc} (\om^{ab} \westar V^c + V^c \westar \om^{ab})
          -i \om \westar v - iv \westar \om
     \ena
 \noi The torsion equation $T= i \psi \westar \psibar $ (\ref{torsioneqD3}) yields:
   \eq
    T^a= {i \over 2} Tr(\psi \westar \psibar \ga^a),~~~t={1 \over 2} Tr (\psi \westar \psibar )
    \en

\subsection{Bianchi identities}

{}From their definition, the curvatures $R,\rho$ and the torsion $T$ satisfy the identities
  \eqa
  & & dR = -R \westar \Om + \Om \westar R \\
  & & d\rho = -R \westar \psi + \Om \westar \rho \label{BianchirhoD3}\\
  & & dT = \Om \westar T -T \westar \Om -R \westar V + V \westar R
   \ena
   \noi The terms on right-hand sides with the spin connection $\Om$ reconstruct
   covariant derivatives on curvatures and torsion, so that the
   identities take the form
   \eq
   DR=0,~~~D\rho=-R \westar \psi,~~~DT=-R \westar V+ V \westar R
   \en

\subsection{Invariances }

The  action (\ref{actionD3}) is invariant under:
\sk

\noi i) {\bf Diffeomorphisms}:

\noi generated by the usual Lie derivative.  Indeed the action is the
 integral of a 3-form on a
3-manifold\footnote{ In order to show that the integrand is a globally defined 3-form we need
to assume that the vielbein one-form $V^a$ is globally defined (and therefore that the
manifold is parallelizable), the twisted exterior product  being globally defined
(because the twist is globally defined). If this is the case, then due to the local
$SO(1,2)\times U(1)$ invariance (see point ii) below) the action is
independent of the vielbein used. On the other hand, if the vielbein $V^a$
is only locally defined  in open coverings of the manifold, then we
cannot construct a global 3-form, since the local
$SO(1,2)\times U(1)$ invariance holds only under integration.},
   \eq
    \int \Lcal_v ({\rm \mbox{3-form}}) = \int (i_v d + d i_v) ({\rm \mbox{3-form}})
    = \int d (i_v ({\rm \mbox{3-form}}))= {\rm  boundary~ term}
    \en
    since $d({\rm \mbox{3-form}}) =0$ on a 3-dimensional manifold.
We have constructed a geometric lagrangian where
    the fields are exterior forms and the $\star$-product is given by the
    Lie derivative action of the twist on forms.   The
    twist $\cal F$ in general is not invariant under the diffeomorphism
    ${\cal L}_v$.  However we can consider the $\star$-diffeomorphisms
    of ref. \cite{Wessgroup} (see also \cite{book}, section 8.2.4),
  generated by the $\star$-Lie derivative. This latter
  acts trivially on the twist $\cal F$ but satisfies a deformed Leibniz
  rule. $\star$-Lie derivatives generate infinitesimal noncommutative
  diffeomorphisms and leave invariant the action and the twist. They are
  noncommutative symmetries of our action.

Finally in our geometric action no coordinate indices $\mu,\nu$ appear, and this
implies invariance of the action under (undeformed) general coordinate
transformations\footnote{General coordinate transformations are
  diffeomorphisms of an open coordinate neighbourhood  of the manifold, not of the
  whole manifold.}.
Otherwise stated every contravariant tensor index $^\mu$ is contracted with the
corresponding covariant tensor index $_\mu$, for example
$X_a=X^\mu_a\partial_\mu$ and $V^a=V^a_\mu dx^\mu$.
\sk
\noi ii)  {\bf Local $SO(1,2) \times U(1) \approx U(1,1)$} variations:
  \eq
\de_\epsilon V = -V \star \epsilon + \epsilon \star V,
~~~\de_\epsilon \Om = d\epsilon - \Om \star \epsilon+ \epsilon
\star \Om,~~~~ \de_\epsilon \psi = \epsilon \star \psi,
~~~\de_\epsilon \psibar = -\psibar \star \epsilon
\label{stargaugeD3}
\en
\noi with
\eq
 \epsilon = \unquarto \epsi^{ab} \ga_{ab} + i \epsi 1
  \en
  \noi satisfying the hermiticity condition:
  \eq
    \ga_0 \epsilon \ga_0 =
 \epsilon^\dagger
   \en
   \noi This condition implies reality of the
component gauge parameters $\epsi^{ab}$, $\epsi$.

  The invariance of (\ref{actionD3}) can be easily checked noting that
   \eq
  \de_\epsilon R = - R \star \epsilon+ \epsilon \star R, ~~~\de_\epsilon \rho = \epsilon \star \rho,~~~\de_\epsilon (\rho \westar \psibar) = - \rho \westar \psibar \star \epsilon + \epsilon \star \rho \westar \psibar
   \en
   \noi and using the cyclicity of the trace $Tr$  and the graded cyclicity of the integral.
   \sk
  \noi  iii) {\bf Local N=1 $\star$-supersymmetry} variations:
       \eq
     \de_\epsilon V = i(\epsilon \star \psibar - \psi \star \epsilonbar),  ~~~ \de_\epsilon \psi = d\epsilon - \Om \star \epsilon \label{susyvarD3}
     \en
    \noi where now $\epsilon$ is a spinorial parameter. Notice that $\Om$ is not varied: we are working in 1.5 - order formalism, i.e.  we are considering $\Om$ as already satisfying its own equation of motion (\ref{torsioneqD3}). Then the  variation of the action due to the supersymmetry variation of $\Om$ vanishes, since it is proportional to the $\Om$ field equation.
  The variations (\ref{susyvarD3}) imply:
     \eq
     \de_\epsilon \psibar = d \epsilonbar + \epsilonbar \star \Om, ~~~ \de_\epsilon \rho = - R \star \epsilon, ~~~
     \de_\epsilon \rhobar = \epsilonbar \star R
     \en
    \noi  The action varies as:
     \eq
     \de_\epsilon S = -2i \int Tr [R \westar (-\psi  \star \epsilonbar + \epsilon \star \psibar) +
       (-R \star \epsilon) \westar \psibar + \rho \westar ( d \epsilonbar + \epsilonbar \star \Om)]
       \en
       \noi After integrating by parts the term with $d \epsilonbar$, using the Bianchi identity for
       $d \rho$ (\ref{BianchirhoD3}) and reordering the $\rho \epsilonbar \Om$ term using the
       cyclicity of $Tr $ and graded cyclicity of the integral, all terms are seen to cancel.
       Thus the action (where $\Om$ is resolved via its equation of motion, i.e. in second order
       formalism)  is invariant under the local $\star$-supersymmetry transformations (\ref{susyvarD3}),
       up to boundary terms.

 \sk
  \noi On the component fields, the $U(1,1)$ transformation rules are:
   \eqa
  & & \de_\epsilon V^a = \unmezzo \epsi^{a}_{~b} \star V^b+ \unmezzo V^b \star \epsi^{a}_{~b}+
   {i \over 4} \epsilon^{abc}
    (v \star \epsi_{bc} -  \epsi_{bc} \star v)+ i( \epsi \star V^a - V^a \star \epsi) \nonumber \\
 & & \de_\epsilon v = - {i \over 4} \epsilon_{abc} (V^a \star \epsi^{bc} -  \epsi^{bc} \star V^a)
     -i(v \star \epsi - \epsi \star v) \nonumber \\
     & & \de_\epsilon \om^{ab} = d \epsi^{ab} + \om^{c[a} \star \epsi_c^{~b]}- \epsi^{c[a} \star \om_c^{~b]}
     -i(\om^{ab} \star \epsi - \epsi \star \om^{ab}) -i(\om \star \epsi^{ab}  - \epsi^{ab} \star \om) \nonumber \\
     & &  \de_\epsilon \om = - d\epsilon - {i \over 8} (\om^{ab} \star \epsi_{ab} -  \epsi^{ab} \star \om_{ab})
     -i(\om \star \epsi - \epsi \star \om) \nonumber \\
      & &\de_\epsilon \psi = \unquarto \epsi^{ab} \ga_{ab} \star \psi + i \epsi \star \psi
    \ena
  \noi and the supersymmetry variations are:
   \eqa
  & & \de_\epsilon V^a = {i \over 2} Tr (\epsilon \star \psibar \ga^a - \psi \star \epsilonbar \ga^a) \nonumber \\
   & & \de_\epsilon v = {1 \over 2} Tr (\epsilon \star \psibar - \psi \star \epsilonbar) \nonumber  \\
     & &  \de_\epsilon \psi = d \epsilon - \unquarto \om^{ab} \ga_{ab} \star \epsilon - i \om \star \epsilon
   \ena
 Finally, it is a straightforward exercise to check that the hermiticity conditions
 on the fields and on the parameters are consistent with the $\star$-gauge and
$\star$-supersymmetry  variations.

\subsection{Commutative limit $\theta \rightarrow 0$}

In the commutative limit the action (\ref{actionD3comp}) reduces to
\eq
 S_{\theta =0} = \int  R^{ab} \we V^c \epsi_{abc} + 4 r \we  v  +  2i \psibar \we \rho \label{actionD3lim1}
 \en
  \noi with
   \eqa
   & & R^{ab} = d\om^{ab}  -  \om^{a}_{~c} \we \om^{cb} , ~~~ r = d\om\\
   & & \rho = d \psi - \unquarto \om^{ab} \ga_{ab} \we \psi - i \om \we \psi
   \ena
   The $\theta = 0$ field equations imply, as in the noncommutative case, that
   all curvatures $R^{ab}, r, \rho$ vanish. The $\theta = 0$ torsion constraints become:
    \eq
    dV^a -\om^{a}_{~b} \we V^b = {i \over 2} \psibar \ga^a \we \psi,~~~dv =  {1 \over 2} \psibar \we \psi
    \label{torsionconstraints}\en

   The term $r \we v = d\om \we v$ in the action  (\ref{actionD3lim1}) can be integrated by parts.
   Using now the second torsion constraint $dv$ can be substituted by $(1/2)(\psibar \we \psi)$, and
   the whole term exactly cancels the $\psibar \om \psi$ term coming from the third term in
    (\ref{actionD3lim1}). Thus the $\theta=0$ action becomes
    \eq
    S_{\theta =0} = \int  R^{ab} \we V^c \epsi_{abc} +   2 i \psibar \we (d\psi -\unquarto  \om^{ab} \ga_{ab}
      \we  \psi ) \label{actionD3lim2}
    \en
   \noi and does not contain any more the fields $\om$ and $v$. In fact it coincides with
   the usual $D=3$ pure supergravity action, involving only the dreibein $V^a$ and the
   gravitino $\psi$. One can at this point use also the first torsion constraint to express $\om^{ab}$
   in terms of the dreibein, retrieving the
   classical action in second order formalism.
   \sk
   \noi{\bf Note:} the second torsion constraint in
   (\ref{torsionconstraints}) implies that $\psibar \we \psi$
   must be closed, which is true on-shell since $d (\psibar \we \psi) =
   \rhobar \we \psi - \psibar \we \rho$.

   \section{Classical  $D=4$, $N=1$ supergravity }

   The $D=4$, $N=1$ simple supergravity action can be written in index-free notation as follows:

   \eq
  S = \int Tr \left[ i R(\Om) \we V \we V \ga_5 -2
( \rho \we \psibar+ \psi \we \rhobar )\we V \ga_5 \right]
\label{actionD4classical}
\en

\noi The fundamental fields are the 1-forms $\Om$  (spin
connection), $V$ (vielbein) and gravitino $\psi$.
The curvature 2-form $R$ and the gravitino curvature $\rho$  are defined by
 \eq
 R= d\Om - \Om \we
\Om, ~~~~~\rho \equiv  D\psi = d\psi - \Om \psi,~~ \rhobar =
D\psibar = d\psibar - \psibar \we \Om
\en
\noi with \eq \Om = {1 \over 4} \om^{ab} \ga_{ab}, ~~~~~V = V^a
\ga_a ~~~~~~
\en
\noi and thus are $4 \times 4$ matrices with spinor indices.
See Appendix C for $D=4$ gamma matrix conventions and useful relations. The Dirac conjugate is
defined as usual: $\psibar = \psi^\dagger \ga_0$. Then also $ \rho \we \psibar$ and $ \psi \we \rhobar$
are matrices in the spinor representation, and the trace  $Tr$ is taken on this representation.
The gravitino field satisfies the Majorana condition:
\eq
 \psi^\dagger \ga_0= \psi^T C
  \en
 \noi  where  $C$ is the $D=4$ charge conjugation matrix, antisymmetric and squaring to $-1$.

Using the $D=4$ gamma matrix  trace identity:
\eq
 Tr (\ga_{ab} \ga_c \ga_d \ga_5) = -4 i \epsi_{abcd}
  \en
 \noi leads to the usual supergravity action in terms of the component fields $V^a$, $\om^{ab}$ :
   \eq
   S = \int  R^{ab} \we V^c \we V^d   \epsi_{abcd}  -4 \psibar \we \ga_5  \ga_a \rho \we V^a
   \label{actionD4comp}
   \en
\noi with
\eq
  R \equiv {1\over4} R^{ab} \ga_{ab},~~~R^{ab} = d\om^{ab}  - \om^{a}_{~c} \we \om^{cb}
\en
 \noi We have also used
 \eq
  \rhobar \ga_5 \ga_a \psi =   \psibar \ga_5 \ga_a \rho
  \en
  \noi due to $\psi$ and $\rho$ being Majorana spinors \footnote{Then the two
  addends in the fermionic part of
  the action (\ref{actionD4classical}) are equal, so that we could have used only one
  of them, with factor $-4$.
  However in the noncommutative extension both will be necessary.}.

  \subsection{Field equations and Bianchi identities}

  Using the cyclicity of the $Tr$ in the action (\ref{actionD4classical}), the variation on $V$, $\Om$ and $\psi$ yield respectively the Einstein equation, the torsion equation and the gravitino equation in index-free form:

   \eq
 Tr[\ga_{a} \ga_5 (- i V \we R  - i R \we V +  2 (\rho \we \psibar + \psi \we \rhobar) ] =0
 \en
 \eq
  Tr[\ga_{ab} \ga_5 (i T \we V - i V \we T +2 \psi \we \psibar \we  V  -2 V \we  \psi \we \psibar)] =0
   \label{torsioneqD4}
  \en
  \eq
  V  \we D\psi =0
  \en
  \noi where the torsion $T = T^a \ga_a$ is defined as:
  \eq
  T \equiv dV -\Om \we V - V \we \Om
   \en
   \noi The solution of the torsion equation (\ref{torsioneqD4}) is given by:
    \eq
     T = i[\psi \we \psibar ,\ga_5]\ga_5 = i  \psi \we \psibar- i \ga_5 \psi \we \psibar
     \ga_5  \label{torsionD4}
     \en
   \noi Upon use of the Fierz identity for Majorana spinor one-forms:
   \eq
    \psi \we \psibar = \unquarto \ga_a \psibar \ga^a \we \psi - {1 \over 8} \ga_{ab} \psibar \ga^{ab} \we \psi
    \en
    \noi the torsion is seen to satisfy the familiar condition
    \eq
    T \equiv T^a \ga_a= {i \over 2} \psibar \ga^a \we \psi \ga_a
    \en
   \noi Finally, the Bianchi identities for the curvatures and the torsion are:
    \eqa
    & & dR = -R \we \Om + \Om \we R  \\
    & & d\rho = -R \we \psi + \Om \we \rho, ~~~ d \rhobar = \psibar\we  R -\rhobar \we \Om \label{BianchirhoD4}\\
    & & dT = - R \we V + \Om \we T - T \we \Om + V \we R
    \ena
    \noi The terms with the spin connection $\Om$ reconstruct covariant derivatives of the curvatures and the
     torsion.

\subsection{Invariances}

We know that the classical supergravity action
(\ref{actionD4comp}) is invariant under general coordinate
transformations, under local Lorentz rotations and under local
supersymmetry transformations. It is of interest to write the
transformation rules of the fields in the index-free notation, so
as to verify the invariances directly on the index-free action
(\ref{actionD4classical}). \sk \noi {\bf Local Lorentz rotations}
 \eq
\de_\epsilon V = -[V,\epsilon ] , ~~~\de_\epsilon \Om = d\epsilon
- [\Om,\epsilon],~~~~ \de_\epsilon \psi = \epsilon \psi,
~~~\de_\epsilon \psibar = -\psibar \epsilon
\label{LorentzD4classical}
\en
\noi with
\eq
 \epsilon = {1\over 4} \epsi^{ab} \ga_{ab}
  \en
  The invariance can be directly checked on the action (\ref{actionD4classical}) noting that
   \eq
  \de_\epsilon R = -[ R, \epsilon],  ~~~\de_\epsilon D\psi = \epsilon D\psi,~~~\de_\epsilon  D\psibar
  = - (D\psibar) \epsilon
   \en
   \noi using the cyclicity of the trace $Tr$ (on spinor indices) and the fact that $\epsilon$ commutes with
   $\ga_5$. The Lorentz rotations close on the Lie algebra:
   \eq
   [\de_{\epsilon_1},\de_{\epsilon_2}] = \de_{[\epsilon_2,\epsilon_1]}
   \en

  \sk
  \noi {\bf Local supersymmetry}
  \sk
  \noi The supersymmetry variations are:
  \eq
     \de_\epsilon V = i [\epsilon \psibar - \psi  \epsilonbar,  \ga_5] \ga_5,~~~ \de_\epsilon \psi =
     D \epsilon \equiv d\epsilon - \Om  \epsilon \label{susyvarD4classical}
     \en
    \noi where now $\epsilon$ is a spinorial parameter (satisfying the Majorana condition).
     Notice that again $\Om$ is not varied since we work in 1.5 - order formalism, i.e. $\Om$ satisfies
      its own equation of motion (\ref{torsioneqD4}).

      The commutator of $\epsilon \psibar - \psi  \epsilonbar$
      with $\ga_5$ in the
    supersymmetry variation of $V$ eliminates the terms even in $\ga_a$
     in the Fierz expansion of
     two generic anticommuting spinors (see Appendix C). Moreover, since $\epsilon$ and
     $\psi$ are Majorana spinors,  the combination
     $\epsilon \psibar - \psi  \epsilonbar$ ensures that only the
     $\ga_a$ component survives. Then (\ref{susyvarD4classical})
       reproduce the usual supersymmetry
    variations (see below).

  The variations (\ref{susyvarD4classical}) imply:
     \eq
     \de_\epsilon \psibar = D \epsilonbar \equiv d \epsilonbar + \epsilonbar  \Om, ~~~ \de_\epsilon \rho = - R  \epsilon, ~~~
     \de_\epsilon \rhobar =  \epsilonbar  R
     \en
     Then the action varies as:
     \eqa
     & &  \de_\epsilon S = \int 2 ~Tr[ R \we ( \psi   \epsilonbar - \epsilon \psibar) \we V \ga_5 +   R \we V \we
      ( \psi   \epsilonbar - \epsilon \psibar) \ga_5 ]- \nonumber \\
       & & - 2 ~Tr [\Big(- R \epsilon \we \psibar \we V + \rho \we (d\epsilonbar + \epsilonbar \Om) \we V
       + (d\epsilon -\Om \epsilon) \we \rhobar \we V + \psi \we \epsilonbar R \we V \Big)\ga_5]\nonumber\\
       & & +2i ~Tr [(\rho \we \psibar + \psi \we \rhobar) ( \psi   \epsilonbar - \epsilon \psibar) \ga_5-
       (\rho \we \psibar + \psi \we \rhobar) \ga_5 ( \psi   \epsilonbar - \epsilon \psibar)]
       \ena

            \noi After integrating by parts the terms with $d \epsilon$ and $d \epsilonbar$, and
            using the Bianchi identity (\ref{BianchirhoD4})  for
       $d \rho$  the variation becomes:
        \eqa
         & & \de_\epsilon S= \int 2~ Tr[ R \we ( \psi   \epsilonbar - \epsilon \psibar) \we V \ga_5 +   R \we V \we
      ( \psi   \epsilonbar - \epsilon \psibar) \ga_5 ]- \nonumber \\
        & &- 2 ~Tr[\Big( - R \epsilon \we \psibar \we V+ \rho \we \epsilonbar \Om \we V
          - \Om \epsilon \we \rhobar \we V +  \psi \we \epsilonbar R \we V + \nonumber\\
        & &  + (R \we \psi - \Om \we \rho) \epsilonbar \we V
            - \rho \epsilonbar \we (T + \Om \we V + V \we \Om) -\nonumber\\
            & & - \epsilon (- \rhobar \we \Om + \psibar \we \rho) \we V -
             \epsilon \rhobar \we (T + \Om \we V + V \we
             \Om)\Big)\ga_5]+
               \nonumber \\
            & &  +2i ~Tr [(\rho \we \psibar + \psi \we \rhobar) ( \psi   \epsilonbar - \epsilon \psibar) \ga_5-
       (\rho \we \psibar + \psi \we \rhobar) \ga_5 ( \psi   \epsilonbar - \epsilon \psibar)]
            \ena

        \noi where we have substituted $dV$ by $T + \Om \we V + V \we \Om$ (torsion definition).
          Using now the cyclicity of $Tr$ , and the fact that $\ga_5$ anticommutes with $V$ and commutes with $\Om$,
          all terms can be easily checked to cancel, except those containing
          the torsion  $T$ and the last line (four-fermion terms).

          Once we make use of the torsion equation ((\ref{torsionD4}) to express $T$ in terms of  gravitino
          fields, the variation reduces to:
           \eqa
            & & \de_\epsilon S= 2 i \int Tr [  \rho \epsilonbar \we (\psi \we \psibar \ga_5 - \ga_5 \psi \we \psibar) +
              \epsilon \rhobar \we (\psi \we \psibar \ga_5 - \ga_5 \psi \we \psibar) \nonumber \\
              & &  ~ +(\rho \we \psibar + \psi \we \rhobar)\we ( \psi   \epsilonbar - \epsilon \psibar) \ga_5-
       (\rho \we \psibar + \psi \we \rhobar)\we \ga_5 ( \psi   \epsilonbar - \epsilon \psibar)]
            \ena
   \noi Finally, carrying out the trace on spinor indices results in

             \eqa
            & & \de_\epsilon S=  2 i \int ( \psibar \epsilon- \epsilonbar \psi) \we ( \psibar \ga_5 \we  \rho - \rhobar \ga_5
               \we \psi)
             + (\psibar \we \rho - \rhobar \we \psi ) \we (\psibar \ga_5 \epsilon - \epsilonbar \ga_5 \psi) \nonumber \\
           & &  + (\epsilonbar \rho - \rhobar \epsilon) \we (\psibar \ga_5 \we \psi) + ( \rhobar \ga_5 \epsilon -
               \epsilonbar \ga_5 \rho) \we (\psibar \we \psi) \label{Svariationclass}
             \ena
            \noi Each factor between parentheses vanishes, due to all spinors being Majorana spinors.
            This proves the invariance of the classical supergravity action under the local supersymmetry
            variations (\ref{susyvarD4classical}).

 \sk
  \noi On the component fields, the Lorentz transformations  (\ref{LorentzD4classical}) read:
   \eqa
  & & \de_\epsilon V^a = \epsi^{a}_{~b} V^b \nonumber \\
     & & \de_\epsilon \om^{ab} = d \epsi^{ab} +  \epsi^{ac} \om_c^{~b} - \epsi^{bc}  \om_c^{~a}
     \nonumber \\
     & &\de_\epsilon \psi = \unquarto \epsi^{ab} \ga_{ab} \psi
    \ena
  \noi and the supersymmetry variations (\ref{susyvarD4classical}) become:
   \eqa
  & & \de_\epsilon V^a = i \epsilonbar \ga^a \psi  \nonumber \\
        & &  \de_\epsilon \psi = d \epsilon - \unquarto \om^{ab} \ga_{ab} \epsilon
   \ena

   \section{Noncommutative $D=4$, $N=1$  supergravity }

   \subsection{Action and $GL(2,C)$ $\star$-gauge symmetry}

A noncommutative generalization of the $D=4$, $N=1$ simple supergravity action is obtained
by replacing exterior products by $\star$-exterior products in (\ref{actionD4classical}):

\eq S = \int Tr \left[ i {R}(\Om) \westar V \westar V \ga_5+ 2
( \rho \westar \psibar+ \psi \westar \rhobar )\westar V \ga_5 \right]
\label{actionD4}
\en

\noi where the curvature 2-form $R$ and the gravitino curvature $\rho$ are
defined as:
 \eq
  {R}= d\Om - \Om \westar \Om, ~~~~~\rho \equiv D\psi = d\psi
- \Om \star \psi
\en

Almost all formulae of the commutative case continue to hold, with ordinary products replaced by
 $\star$-products and $\star$-exterior products. However, the expansion of the fundamental fields on the Dirac basis of gamma matrices must now include new contributions; more precisely  the spin connection contains all even gamma matrices
 and  the vielbein contains all
 odd gamma matrices:
 \eq
  \Om = {1 \over 4} \om^{ab} \ga_{ab} + i \om 1 + \omtilde \ga_5, ~~~~~V = V^a
\ga_a + \Vtilde^a \ga_a \ga_5  ~~~~~~
\en
\noi The one-forms $\Om$ and $V$ are thus also $4 \times 4$ matrices with spinor indices. Similarly for the curvature :
 \eq
 {R}= {1\over 4} R^{ab} \ga_{ab} + i r 1 + \rtilde \ga_5
  \en
  \noi and for the gauge parameter:
 \eq
 \epsilon = {1\over 4} \epsi^{ab} \ga_{ab} + i \epsi 1 + \epsitilde \ga_5
  \en
  \noi Indeed now the $\star$-gauge variations read:
  \eq
\de_\epsilon V = -V \star \epsilon + \epsilon \star V,
~~~\de_\epsilon \Om = d\epsilon - \Om \star \epsilon+ \epsilon
\star \Om,~~~~ \de_\epsilon \psi = \epsilon \star \psi,
~~~\de_\epsilon \psibar = -\psibar \star \epsilon
\label{stargauge}\en
 \noi and in the variations for $V$ and $\Om$ also anticommutators of gamma matrices appear,
 due to the noncommutativity of the $\star$-product. Since for example the anticommutator
 $\{ \ga_{ab},\ga_{cd} \}$ contains $1$ and $\ga_5$, we see that the corresponding fields
 must be included in the expansion of $\Om$. Similarly, $V$ must contain a $\ga_a \ga_5$ term due
 to $\{ \ga_{ab},\ga_{c} \}$. Finally, the composition law for gauge parameters becomes:
 \eq
   [\de_{\epsilon_1},\de_{\epsilon_2}] = \de_{\epsilon_2 \star \epsilon_1 - \epsilon_1 \star \epsilon_2 }
   \en
   \noi so that $\epsilon$ must contain the $1$ and $\ga_5$ terms, since they appear in the
   composite parameter $\epsilon_2 \star \epsilon_1 - \epsilon_1 \star \epsilon_2$.

   The invariance of the noncommutative action (\ref{actionD4}) under the $\star$-gauge variations is
   demonstrated  in exactly the same way as for the commutative case, noting that
   \eq
  \de_\epsilon R = - R \star \epsilon+ \epsilon \star R, ~~~\de_\epsilon D\psi = \epsilon \star D\psi,~~~\de_\epsilon ((D\psi) \westar \psibar) = - (D\psi) \westar \psibar \star \epsilon + \epsilon \star (D\psi) \westar \psibar
   \en
   \noi and using now, besides the cyclicity of the trace $Tr$  and the fact that $\epsilon$ still commutes with $\ga_5$, also the graded cyclicity of the integral.

\subsection{Local $\star$-supersymmetry }

The $\star$-supersymmetry variations are obtained from the
classical ones using $\star$-products: \eq
     \de_\epsilon V = i [\epsilon \star \psibar - \psi \star \epsilonbar,  \ga_5] \ga_5~~~ \de_\epsilon \psi = d\epsilon - \Om \star  \epsilon \label{susyvarD4}
     \en
    \noi where $\epsilon$ is a spinorial parameter. Under these variations the noncommutative action varies as given in
  (\ref{Svariationclass}), with ordinary products substituted with $\star$-products.
  Indeed the algebra is identical, since $\ga_5$ still anticommutes with $V$ and commutes with $\Om$,
  and we can use the cyclicity of $Tr$ and graded cyclicity of the integral.

  The question is now: does this variation vanish? Classically it vanishes because of the Majorana
  condition on the spinors (gravitino and supersymmetry gauge parameter).
  We recall the noncommutative generalization of the Majorana
  condition, consistent with the $*$-gauge
  transformations  \cite{AC1}:
    \eq
    \psi^c_\theta = \psi_{-\theta}^{},~~~~~~\psi^c \equiv C(\psibar)^T \label{NCMajorana}
     \en
 This condition involves the $\theta$ dependence of the fields
\footnote{The fields
  can be formally expanded in powers of $\theta$: in principle this picture would introduce infinitely many fields, one for each power of $\theta$. However the Seiberg-Witten map \cite{SW,Jurco} can be used to express all fields in terms of the classical one, ending up with a finite number of fields.},
  and is consistent with the $\star$-gauge transformations only if the
 gauge parameter satisfies the charge conjugation condition \cite{AC1}:
  \eq
  C \epsilon^{}_\theta C = \epsilon^{T}_{-\theta} \label{Cconjpar}
   \en
  \noi The NC Majorana condition (\ref{NCMajorana}) is consistent also with
  $\star$-supersymmetry transformations if the supersymmetry parameter is Majorana, and
   the bosonic fields satisfy the charge conjugation conditions
   \eq
   C \Om^{}_\theta C = \Om^T_{-\theta},~~~C V^{}_\theta C = V^T_{-\theta} \label{Cconjfields}
    \en

  Now consider the first term in the supersymmetry variation of the action (for the other three terms the reasoning is identical):
    \eq
     2 i \int ( \psibar \star \epsilon- \epsilonbar \star \psi) \westar ( \psibar \ga_5 \westar  \rho - \rhobar \ga_5
               \westar \psi) \label{term1susyvar}
       \en
       \noi If $\psi$ and $\epsilon$ are noncommutative Majorana fermions, they satisfy the relations:
       \eq
       \psibar \star \epsilon = \epsilonbar_{-\theta}^{} \star_{-\theta}^{}
       \psi_{-\theta}^{},~~~~ \psibar \ga_5 \westar \rho =
      \rhobar_{-\theta}^{} \ga_5  \we_{-\theta}^{} \psi_{-\theta}^{}
      \en
   \noi  and one sees that (\ref{term1susyvar}) does not vanish anymore (although it vanishes in the
   commutative limit).
    Thus the NC Majorana condition does not ensure the local $\star$-supersymmetry invariance of
    the action in (\ref{actionD4}). In fact, the local supersymmetry of the commutative action
    is broken by noncommutativity.
  \sk

   There is another condition that we
   can impose on fermi fields, the Weyl condition, still
   consistent with the $\star$-symmetry structure of the action:
    \eq
    \ga_5 \psi= \psi, ~~~ \ga_5 \epsilon = \epsilon
    \en
    \noi i.e. all fermions are left-handed (so that their Dirac conjugates
    $\psibar$ and $\epsilonbar$ are right-handed). In this case the local
    $\star$-supersymmetry
    variation vanishes because in all the fermion bilinears the $\ga_5$ matrices
    can be omitted, and the product of a right-handed spinor with a left-handed spinor
    vanishes. Thus the noncommutative supergravity action (\ref{actionD4}) with Weyl fermions is
   locally supersymmetric.

   Note that now we cannot impose the charge conjugation relations   (\ref{Cconjfields}) on the bosonic fields
: indeed $\star$-supersymmetry links together these relations with the NC Majorana condition, which is not
compatible in $D=4$ with the Weyl condition (as in the classical case).

   The $\theta \rightarrow 0$ limit  of this chiral noncommutative theory is a complex version of the so-called $D=4$, $N=1$
   Weyl supergravity and is discussed in Section 4.6 below.

     \subsection{Hermiticity conditions and reality of the action}

Hermiticity conditions can be imposed on $V$, $\Om$ and the gauge
parameter $\epsilon$: \eq
 \ga_0 V \ga_0 = V^\dagger,~~~ -\ga_0 \Omega \ga_0 =
 \Omega^\dagger,~~~ -\ga_0 \epsilon \ga_0 =
 \epsilon^\dagger
 \en
 \noi Moreover it is easy to verify that :
 \eq
 \ga_0 [\rho \westar \psibar] \ga_0 = [\psi \westar \rhobar]^\dagger
\en
\noi These conditions are consistent with the $\star$-gauge and
$\star$-supersymmetry variations (both for Majorana and chiral fermions), as
in the commutative case, and can be used to check that the action
(\ref{actionD4}) is real. The hermiticity conditions imply
that  the component fields $V^a$,
$\Vtilde^a$, $\om^{ab}$, $\om$, and $\omtilde$, and
gauge parameters $\epsi^{ab}$, $\epsi$, and $\epsitilde$
 are real fields.

 \subsection{Component analysis}

 Here we list the $\star$-gauge and supersymmetry variations of the component
 fields. In the supersymmetry variations we consider both Majorana and Weyl fermions.

\subsubsection{ $\star$-Gauge variations}

\eqa
 & & \de_\epsilon V^a = \unmezzo (\epsi^{a}_{~b} \star V^b + V^b \star
 \epsi^{a}_{~b}) + {i \over 4} \epsi^{a}_{~bcd} (\Vtilde^{b}
 \star \epsi^{cd} -  \epsi^{cd} \star \Vtilde^{b}) \nonumber \\
& &~~~~~~~~ + \epsi \star V^a - V^a \star \epsi - \epsitilde \star
\Vtilde^a - \Vtilde^a \star \epsitilde\\
 & & \de_\epsilon \Vtilde^a = \unmezzo (\epsi^{a}_{~b} \star \Vtilde^b + \Vtilde^b \star
 \epsi^{a}_{~b}) + {i \over 4} \epsi^{a}_{~bcd} (V^{b}
 \star \epsi^{cd} -  \epsi^{cd} \star V^{b}) \nonumber \\
& &~~~~~~~~ + \epsi \star \Vtilde^a - \Vtilde^a \star \epsi -
\epsitilde \star V^a - V^a \star \epsitilde\\
 & & \de_\epsilon \om^{ab} = \unmezzo (\epsi^a_{~c} \star \om^{cb} -\epsi^b_{~c} \star \om^{ca}
   + \om^{cb} \star  \epsi^a_{~c} - \om^{ca} \star \epsi^b_{~c})
   \nonumber \\
   & & ~~~~~~ + {1 \over 4} (\epsi^{ab} \star \om - \om \star
   \epsi^{ab}) + {i \over 8} \epsi^{ab}_{~~cd} (\epsi^{cd} \star
   \omtilde - \omtilde \star \epsi^{cd}) \nonumber \\
   & & ~~~~~~+ {1\over 4} (\epsi \star \om^{ab} - \om^{ab} \star \epsi)
   + {i \over 8} \epsi^{ab}_{~~cd} (\epsitilde \star
   \om^{cd} - \om^{cd} \star \epsitilde)\\
  & & \de_{\epsilon} \om = {1\over 8} (\om^{ab} \star \epsi_{ab} -
   \epsi_{ab} \star \om^{ab}) + \epsi \star \om - \om \star \epsi
   + \epsitilde \star \omtilde - \omtilde \star \epsitilde\\
     & & \de_{\epsilon} \omtilde = {i \over 16} \epsi_{abcd}
    (\om^{ab} \star \epsi^{cd} - \epsi^{cd} \star \om^{ab}) +
    \epsi \star \omtilde - \omtilde \star \epsi + \epsitilde \star
    \om - \om \star \epsitilde
    \ena

\subsubsection{Supersymmetry variations: Majorana fermions}
  \eqa
& & \de_\epsilon V^a= {i \over 2} Tr [(\epsilon \star \psibar - \psi \star \epsilonbar)\ga^a] \\
& & \de_\epsilon \Vtilde^a=    {i \over 2} Tr [(\epsilon \star \psibar - \psi \star \epsilonbar)\ga^a \ga_5]\\
 & & \de_\epsilon \psi= d \epsilon -  {1 \over 4} \om^{ab} \ga_{ab} \epsilon - (i \om
   + \omtilde \ga_5) \epsilon
   \ena

  \subsubsection{Supersymmetry variations: Weyl fermions}
\eqa
& & \de_\epsilon V^a= \de_\epsilon \Vtilde^a= {i \over 2} Tr [(\epsilon \star \psibar - \psi \star \epsilonbar)\ga^a]  \\
 & & \de_\epsilon \psi= d \epsilon -  {1 \over 4} \om^{ab} \ga_{ab} \epsilon - (i \om
   + \omtilde) \epsilon
   \ena

 \subsubsection{Charge conjugation conditions}

  The charge conjugation relations
 (\ref{Cconjfields}) imply for the component fields:

 \eqa & & V^a_\theta=V^a_{-\theta}, ~~~
\om^{ab}_\theta = \om^{ab}_{-\theta} \\ & & \Vtilde^a_\theta
=-\Vtilde^a_{-\theta}, ~~~ \om^{}_\theta=- \om^{}_{-\theta} ,~~~\omtilde^{}_\theta=
- \omtilde^{}_{-\theta}, \label{cconjonfields}
 \ena
 \noi and
for the gauge parameters:
 \eqa & & \epsi^{ab}_\theta= \epsi^{ab}_{-\theta}
 \\ & &  \epsi^{}_\theta =- \epsi^{}_{-\theta} ,~~~\epsitilde^{}_\theta=-
\epsitilde^{}_{-\theta} \label{cconjonparam}
 \ena

 \subsection{Field equations and Bianchi identities}

  Using the cyclicity of the integral and of the $Tr$ in the action
  (\ref{actionD4}), the variation on $V$, $\Om$ and $\psi$ yield respectively the Einstein equation, the torsion equation and the gravitino equation in index-free form:

   \eq
 Tr[\Ga_{a,a5} (- i V \westar R  - i R \westar V +
  2 (\rho \westar \psibar + \psi \westar \rhobar) ] =0
 \en
 \eq
  Tr[\Ga_{ab,1,5} (i T \westar V - i V \westar T +
  2 \psi \westar \psibar \we  V  - 2 V \westar  \psi \westar \psibar)] =0
   \label{torsioneqD4NC}
  \en
  \eq
  V \westar D\psi-{1 \over 2} T \westar \psi =0
  \en
   \noi where $\Ga_{ab,1,5}$ indicates $\ga_{ab}$, $1$ and $\ga_5$
  (thus there are three distinct equations) and likewise for
  $\Ga_{a,a5}$ (two equations corresponding to $\ga_a$ and $\ga_a
  \ga_5$). The torsion $T = T^a \ga_a+ \Ttilde^a \ga_a \ga_5$ is defined as:
  \eq
  T \equiv dV -\Om \westar V - V \westar \Om
   \en
   \noi The torsion equation can be written as:
   \eq
   [i T \westar V - i V \westar T + 2 \psi \westar \psibar \westar
   V - 2 V \westar \psi \westar \psibar, \ga_5] =0
   \en
   \noi since the anticommutator with $\ga_5$ selects the $\ga_{ab}$, $1$ and
   $\ga_5$ components. This equation can be solved for the
   torsion:
   \eq
    T = i [\psi \westar \psibar, \ga_5] \ga_5= i \psi \westar
    \psibar - i \ga_5 \psi \westar \psibar \ga_5  \label{torsionD4NC}
    \en
    \noi For chiral gravitini:
    \eq
    T = 2i \psi \westar \psibar
    \en

   \noi The Bianchi identities for the curvatures and the torsion are obtained
   from the commutative ones simply by replacing exterior products
   by $\star$-exterior products.

\subsection{Commutative limit}

The nonsupersymmetric NC theory with NC Majorana gravitino, and
charge conjugation conditions (\ref{Cconjfields}), reduces in the
$\theta \rightarrow 0$ limit to the usual $D=4$, $N=1$
supergravity.
 Indeed the charge conjugation conditions on $V$ and $\Om$ imply that
the component fields  $\Vtilde^a$, $\om$, and $\omtilde$ all vanish in the limit $\theta \rightarrow 0$
(see the second line of (\ref{cconjonfields})), and only the classical spin connection
$\om^{ab}$, vierbein $V^a$ and Majorana fermion $\psi$ survive. Similarly the gauge parameters
$\epsi$, and $\epsitilde$ vanish in the commutative limit.

 \sk
 In the chiral case,  the extra vielbein $\Vtilde^a$ cannot vanish in the commutative limit, since its
supersymmetry variation is equal to that of $V^a$. Then one obtains a
commutative limit that is a (locally) supersymmetric version of gravity
with a complex vielbein studied by Chamseddine, or a bigravity-like theory
(in our case a super-bigravity theory). For a discussion on chiral supergravity
see for ex. \cite{Mielke}. A detailed study of this commutative
limit will not be carried out in the present paper.

\subsection{The noncommutative supergravity action in terms of chiral fields}

In the case of chiral fermions, it may be useful to
reexpress the action in terms of chiral bosonic and fermionic
fields. Chiral bosonic fields can be defined in exactly the same
way as chiral fermionic fields, since $V$ and $\Om$ take values in
the spinor representation (they are Clifford algebra valued
fields). Thus we'll denote by $V_\pm$ and $\Om_\pm$ the
projections \eq
 V_\pm = \unmezzo (1 \pm \ga_5) V, ~~~\Om_\pm =  \unmezzo (1 \pm
 \ga_5)\Om
 \en
 Note that the spin connection $\om^{ab}$ contained in $\Om_\pm$
 is then (anti)self-dual.

 The action (\ref{actionD4}) takes the form:

 \eq
 S = \int Tr [i {R}_+ \westar V_+ \westar V_- -i {R}_- \westar V_- \westar
 V_+ + 2 (\rho \westar \psibar + \psi \westar \rhobar) \westar
 V_-]
  \en

 \noi with
 \eq
 {R}_\pm = d \Om_\pm - \Om_\pm \westar \Om_\pm
  \en

The transformation rules and the field equations can all be
rewritten in terms of the chiral fields. For example under
supersymmetry the  ``chiral vielbein"  $V_\pm$ transform as:

\eq
 \de_\epsilon V_+ = 2i(\epsilon \star \psibar - \psi \star
 \epsilonbar),~~~  \de_\epsilon V_- =0
 \en
  Similarly the torsion equation becomes:
  \eq
  T_+ = 2i \psi \westar \psibar,~~~T_-=0
   \en

\section{Conclusions}

The index-free notation, based on Clifford algebra expansion
of the bosonic fields (see for ex. ref.s \cite{Mielke, Chamseddine}),
allows to study invariances with simple algebraic manipulations.
This framework is ideally suited to study noncommutative generalizations
of field theories containing gravity, cf. ref.s
\cite{Chamseddine}, where a complex noncommutative gravity
was proposed. In ref. \cite{AC1} we showed that a NC gravity
could be constructed, with a commutative limit coinciding with the
usual Einstein-Cartan theory. We proved that
 a NC charge conjugation condition on the vierbein and on the
 spin connection yields a real vierbein in the commutative limit.
 The theory was also coupled to
(Majorana) fermion zero-forms (spin 1/2).

In this paper we have constructed noncommutative supergravities
in $D=3$ and $D=4$.  The commutative limit of the $D=3$ locally
supersymmetric theory coincides with pure supergravity (without cosmological term)
in $D=3$. The $D=4$ model is less satisfactory: if we use the NC Majorana
condition for the gravitino, the action is not $\star$-supersymmetric. However
in this case we can impose charge conjugation conditions on the vierbein and
spin connection, so that the commutative limit of the theory
reproduces usual $D=4$, $N=1$ supergravity.

 We recover $\star$-local supersymmetry of the action
when the gravitino is chiral. In this case we cannot impose the
charge conjugation condition on the vierbein (because then
$\star$-supersymmetry requires the NC Majorana condition on the
gravitino), and therefore the commutative limit does not involve
only one real vierbein, but reduces to a chiral $D=4$, $N=1$
supergravity with a complex vierbein.

Note that the $\star$-products deformations considered in this paper are
associated to a very general triangular Drinfeld twist $\cal F$, a particular case being the Groenewold-Moyal $\star$-product.  In our
general framework one
could consider promoting the twist $\cal F$ itself to a dynamical
field, see \cite{ACD} for an example in the flat case.

 \sect{Appendix A: twist differential geometry}

The noncommutative deformation of the gravity theories we constructed
relies on the existence (in the deformation quantization context, see
for ex \cite{book} ) of an associative $\star$-product between
functions and more generally an associative $\westar$ exterior product between forms that
satisfies the following properties:
\sk
\noi
$\bullet~~$ \noi Compatibility with the undeformed exterior differential:
\eq
d(\tau\wedge_\star \tau')=d(\tau)\wedge_\star \tau'=\tau\wedge_\star
d\tau'
\en
$\bullet~~$ Compatibility with the undeformed integral (graded cyclicity property):
        \eq
       \int \tau \westar \tau' =  (-1)^{deg(\tau) deg(\tau')}\int \tau' \westar \tau\label{cycltt'}
       \en
      \noi with $deg(\tau) + deg(\tau')=$D=dimension of the spacetime
      manifold, and where here $\tau$ and $\tau'$ have compact support
      (otherwise stated we require (\ref{cycltt'}) to hold up to
      boundary terms).
\sk
\noi $\bullet~~$ Compatibility with the undeformed complex conjugation:
\eq
       (\tau \westar \tau')^* =   (-1)^{deg(\tau) deg(\tau')} \tau'^* \westar \tau^*
\en
 We describe here a (quite wide) class of twists whose   $\star$-products
 have all these properties.
In this way we have constructed a wide class of noncommutative
deformations of gravity theories. Of course as a particular case we
have the Groenewold-Moyal $\star$-product
\begin{equation}
f\star g = \mu \big{\{} e^{\frac{i}{2}\theta^{\rho\sigma}\partial_\rho \otimes\partial_\sigma}
f\otimes g \big{\}} , \label{MWstar}
\end{equation}
where the map $\mu$  is the usual pointwise
multiplication: $\mu (f \otimes g)= fg$, and $\theta^{\rho\sigma}$ is a constant
antisymmetric matrix.

\sk

\noi{\bf Twist}
\sk
\noi Let $\Xi$ be the linear space of smooth vector fields on a smooth manifold $M$, and $U\Xi$ its
universal enveloping algebra. A twist  ${\cal F} \in U\Xi \otimes U\Xi$
defines the associative twisted product
\begin{eqnarray}
f\star g &=& \mu \big{\{} {\cal F}^{-1} f\otimes g \big{\}}
\end{eqnarray}
\noi  where the map $\mu$  is the usual pointwise
multiplication: $\mu (f \otimes g)= fg$. The product associativity relies on the defining properties of the twist \cite{Wessgroup,book,Aschieri}.
Using the standard notation
\eq {\cal F}
\equiv \rf^\alpha \otimes \rf_\alpha,   ~~~
  {\cal F}^{-1}
\equiv \overline{\rf}^\alpha \otimes \overline{\rf}_\alpha
\en
 \noi (sum over $\alpha$ understood) where $\rf^\al, \rf_\al, \overline{\rf}^\alpha , \overline{\rf}_\alpha$ are elements of  $U\Xi$, the
 $\star$-product is expressed in terms of ordinary products as:
  \eq
  f \star g =  \overline{\rf}^\alpha (f)  \overline{\rf}_\alpha (g)
   \en
   \noi Many explicit examples of twist are provided by the so-called abelian twists:
\eq
{\cal F}= e^{-\frac{i}{2}\theta^{ab}X_a \otimes X_b} \label{Abeliantwist}
\en
where $\{X_a\}$ is a set of mutually commuting vector fields globally
defined on the manifold\footnote{
We actually need only the twist $\cal F$ to be globally defined, not
necessarily the single vector fields $X_a$.  An explicit example of
this latter kind is given by the  twist
(\ref{Abeliantwist}), that in an open neighbourhood with
coordinates $t,x,y,z$ is defined by the commuting vector fields
$X_1=f(x,z){\partial\over \partial x}$,
$X_2=h(y,z){\partial\over \partial y}$, where $f(x,z)$ is
a function of only the $x$ and $z$ variables and has
compact  support, and similarly $h(y,z)$.  This twist is globally defined on the whole manifold by
  requiring it to be the identity $1\otimes 1$ outside the $\{x^a\}$
  coordinate neighbourhood. The corresponding $\star$-product,
  defined on the whole spacetime manifold, is noncommutative only
  inside this neighbourhood.
}, and $\theta^{ab}$ is a constant
antisymmetric matrix. The corresponding $\star$-product is in general
position dependent because the vector fields $X_a$ are in general
$x$-dependent. In the special case that there exists a
global coordinate system on the manifold we can consider the
vector fields $X_a={\partial \over \partial x^a}$. In this instance we have
the Moyal twist, cf. (\ref{MWstar}):
  \eq
   {\cal F}^{-1}=  e^{\frac{i}{2}\theta^{\rho\sigma}\partial_\rho \otimes\partial_\sigma} \label{Mtwist}
   \en
  \noi {\bf Deformed exterior product}
    \sk

   \noi The deformed exterior product between forms is defined as
   \eq
   \tau \westar \tau' \equiv \fbaralup  (\tau) \we \fbaraldo (\tau') \label{defwestar}
       \en
       \noi where $ \fbaralup$ and $\fbaraldo$ act on forms via the Lie derivatives
       ${\cal L}_{ \fbaralup} $,  ${\cal L}_{ \fbaraldo} $
       (Lie derivatives along products $uv \cdots$ of elements of  $\Xi$ are defined
       simply by $ {\cal L}_{uv \cdots} \equiv {\cal L}_u {\cal L}_v \cdots$).
     This product is associative, and in particular satisfies:
   \eq
    \tau \westar h \star \tau' = \tau \star h \westar \tau',~~~h \star (\tau \westar \tau') = (h \star \tau) \westar \tau',~~~
    (\tau \westar \tau') \star h = \tau \westar (\tau' \star h)
    \en
    \noi where $h$ is a $0$-form,  i.e. a function  belonging to $Fun(M)$, the $\star$-product
    between functions and one-forms being just a particular case of  (\ref{defwestar}):
     \eq
     h \star \tau = \fbaralup (h) \fbaraldo  (\tau), ~~~\tau \star h = \fbaralup (\tau) \fbaraldo (h)
     \en

\noi {\bf Exterior derivative}
        \sk
         \noi The exterior derivative satisfies the usual (graded) Leibniz rule,
         since it commutes with the Lie derivative:
        \eqa
        & & d (f \star g) = df \star g + f \star dg \\
        & & d(\tau \westar \tau') = d\tau \westar \tau'  + (-1)^{deg(\tau)} ~\tau \westar d\tau'
        \ena

\sk

       \noi {\bf Integration: graded cyclicity} \nopagebreak
        \sk
        \noi If we consider an abelian twist (\ref{Abeliantwist})
        given by globally defined commuting vector fields $X_a$,
        then the usual integral is cyclic under the $\star$-exterior
        products of forms, i.e., up to boundary terms,
        \eq
       \int \tau \westar \tau' =  (-1)^{deg(\tau) deg(\tau')}\int \tau' \westar \tau
       \en
      \noi with $deg(\tau) + deg(\tau')=$D=dimension of the spacetime
      manifold. In fact we have
\eq       \int \tau \westar \tau' =    \int \tau \wedge \tau'=
(-1)^{deg(\tau) deg(\tau')}\int \tau' \wedge \tau=
(-1)^{deg(\tau) deg(\tau')}\int \tau' \westar \tau
\en
For example at first order in $\theta$,
\eq
\int \tau \westar \tau' =    \int \tau \wedge \tau'-{i\over
 2}\theta^{ab}\int{\cal L}_{X_a}(\tau\wedge {\cal L}_{X_b}\tau')
=
\int \tau \wedge \tau'-{i\over
 2}\theta^{ab}\int d {i}_{X_a}(\tau\wedge {\cal L}_{X_b}\tau')
\en
where we used the Cartan formula ${\cal L}_{X_a}=di_{X_a}+i_{X_a}d$.
\sk
More generally if the twist $\cal F$ satisfies the condition
$S(\fbaralup)\fbaraldo=1$,
where the antipode $S$ is defined on vector fields
        as  $S(v)=-v$  and is extended to the whole universal
        enveloping algebra $U\Xi$ linearly and antimultiplicatively,
        $S(uv)=S(v)S(u)$, then a similar argument proves the graded
        cyclicity of the integral\footnote{Proof: using Sweedler's coproduct notation
          (cf. \cite{Wessgroup})) we have
\eqa\tau\wedge_\star \tau'&=&   \fbaralup  (\tau) \we \fbaraldo (\tau') =
 \fbaralup_1  (\tau \we  S(\fbaralup_2)\fbaraldo (\tau'))=
\tau \we  S(\fbaralup)\fbaraldo (\tau')+ {\fbaralup}'_1  (\tau \we
S({\fbaralup}'_2)\fbaraldo (\tau'))\nonumber\\ &=&\tau \we  \tau'+
{\rm total~derivative}\nonumber \ena \noi where $\Delta \fbaralup
\equiv \fbaralup_1 \otimes \fbaralup_2 \equiv 1 \otimes \fbaralup
+ {\fbaralup}'_1 \otimes {\fbaralup}'_2$, and
 in the last equality we
observe that each ${\fbaralup}'_1$ contains at least one vector
field. Thus use of Cartan's formula implies that the second addend
is a total derivative.}. \sk
        \noi {\bf Complex conjugation}
    \sk
        \noi If we choose real fields $X_a$ in the definition of the
        twist (\ref{Abeliantwist}),  it is immediate to verify that:
        \eq
        (f \star g)^* = g^* \star f^*\label{starfg*}
        \en
        \eq
        (\tau \westar \tau')^* =   (-1)^{deg(\tau) deg(\tau')} \tau'^* \westar \tau^*\label{startt*}
        \en
        since sending $i$ into $-i$ in the twist (\ref{Mtwist}) amounts to send $\theta^{ab}$ into
        $-\theta^{ab} = \theta^{ba}$, i.e. to exchange the
        order of the factors in the $\star$-product.
\sk
 More in general
        we can consider twists $\cal F$ that satisfy the reality condition
        (cf. Section 8 in \cite{Wessgroup} )
        ${\fbaralup}^* \otimes {\fbaraldo}^*=S(\fbaraldo) \otimes S
        (\fbaralup)$. The $\star$-products associated to these
        twists satisfy properties (\ref{starfg*}), (\ref{startt*}).
        \sk
        \sk

\sect{Appendix B : gamma matrices in $D=3$}

We summarize in this Appendix our gamma matrix conventions in $D=3$.

\eq
 \ga_0 =
\left(
\begin{array}{cc}
  i &  0    \\
  0&  -i
\end{array}
\right),~~~\ga_1=
\left(
\begin{array}{cc}
  0&   1  \\
  1&     0
\end{array}
\right)
,~~~\ga_2=
\left(
\begin{array}{cc}
  0&   -i \\
  i&     0
\end{array}
\right)
\en

\eqa
& & \eta_{ab} =(-1,1,1),~~~\{\ga_a,\ga_b\}=2 \eta_{ab},~~~[\ga_a,\ga_b]=2 \ga_{ab}= -2 \epsi_{abc} \ga^c, \\
& & \epsi_{012} = - \epsi^{012}=1, \\
& & \ga_a^\dagger = \ga_0 \ga_a \ga_0
\ena

\subsection{Useful identities}

\eqa
 & &\ga_a\ga_b= \ga_{ab}+\eta_{ab}= - \epsi_{abc} \ga^c + \eta_{ab}\\
 & &\ga_{ab} \ga_c=\eta_{bc} \ga_a - \eta_{ac} \ga_b -\epsi_{abc}\\
 & &\ga_c \ga_{ab} = \eta_{ac} \ga_b - \eta_{bc} \ga_a -\epsi_{abc}\\
 & &\ga_a\ga_b\ga_c= \eta_{ab}\ga_c + \eta_{bc} \ga_a - \eta_{ac} \ga_b - \epsi_{abc}\\
 & &\ga^{ab} \ga_{cd} = - 4 \de^{[a}_{[c} \ga^{b]}_{~~d]} - 2 \de^{ab}_{cd}
  \ena
 \noi where $\de^{ab}_{cd}
 = \unmezzo (\de^a_c \de^b_d - \de^a_d \de^b_c)$, and index antisymmetrizations in square brackets have weight 1.

\sect{Appendix C : gamma matrices in $D=4$}

We summarize in this Appendix our gamma matrix conventions in $D=4$.

\eqa
& & \eta_{ab} =(1,-1,-1,-1),~~~\{\ga_a,\ga_b\}=2 \eta_{ab},~~~[\ga_a,\ga_b]=2 \ga_{ab}, \\
& & \ga_5 \equiv i \ga_0\ga_1\ga_2\ga_3,~~~\ga_5 \ga_5 = 1,~~~\epsi_{0123} = - \epsi^{0123}=1, \\
& & \ga_a^\dagger = \ga_0 \ga_a \ga_0, ~~~\ga_5^\dagger = \ga_5 \\
& & \ga_a^T = - C \ga_a C^{-1},~~~\ga_5^T = C \ga_5 C^{-1}, ~~~C^2 =-1,~~~C^T =-C
\ena

\subsection{Useful identities}

\eqa
 & &\ga_a\ga_b= \ga_{ab}+\eta_{ab}\\
 & & \ga_{ab} \ga_5 = {i \over 2} \epsilon_{abcd} \ga^{cd}\\
 & &\ga_{ab} \ga_c=\eta_{bc} \ga_a - \eta_{ac} \ga_b -i \epsi_{abcd}\ga_5 \ga^d\\
 & &\ga_c \ga_{ab} = \eta_{ac} \ga_b - \eta_{bc} \ga_a -i \epsi_{abcd}\ga_5 \ga^d\\
 & &\ga_a\ga_b\ga_c= \eta_{ab}\ga_c + \eta_{bc} \ga_a - \eta_{ac} \ga_b -i \epsi_{abcd}\ga_5 \ga^d\\
 & &\ga^{ab} \ga_{cd} = -i \epsi^{ab}_{~~cd}\ga_5 - 4 \de^{[a}_{[c} \ga^{b]}_{~~d]} - 2 \de^{ab}_{cd}
 \ena

 \subsection{Charge conjugation and Majorana condition}

 \eqa
 & &   {\rm Dirac~ conjugate~~} \psibar \equiv \psi^\dagger
 \ga_0\\
 & &  {\rm Charge~ conjugate~spinor~~} \psi^c = C (\psibar)^T  \\
 & & {\rm Majorana~ spinor~~} \psi^c = \psi~~\Rightarrow \psibar =
 \psi^T C
 \ena

\subsection{Fierz identities for two spinor one-forms}

\eq
 \psi \we \chibar = \unquarto [ (\chibar \we \psi) 1 + (\chibar \ga_5 \we \psi) \ga_5 + (\chibar \ga^a \we \psi) \ga_a + (\chibar \ga^a \ga_5 \we \psi) \ga_a \ga_5  - \unmezzo (\chibar \ga^{ab} \we \psi) \ga_{ab}]
 \en

\noi {\bf Noncommutative Fierz identities}
\eqa
& & \psi \westar \chibar = \unquarto [ Tr (\psi \westar \chibar) 1 + Tr (\psi \ga_5 \westar \chibar) \ga_5 + Tr (\psi \ga^a \westar \chibar) \ga_a + \nonumber \\
& & ~~~~~~~~~~~~~~~Tr (\psi \ga^a \ga_5 \westar \chibar) \ga_a \ga_5  - \unmezzo Tr (\psi \ga^{ab} \westar \chibar) \ga_{ab}]
 \ena

 \vfill\eject


\begin{thebibliography}{99}

 \bibitem{Castellaniqgrav}
  L.~Castellani,
  {\it Differential calculus on $ISO_q(N)$, quantum Poincar\'e algebra and
  q-gravity,}
  Commun.\ Math.\ Phys.\  {\bf 171}, 383 (1995)
  [hep-th/9312179];
   { \it The Lagrangian of q-Poincar\'e gravity,}
  Phys.\ Lett.\  B {\bf 327}, 22 (1994)
  [hep-th/9402033].
  %%CITATION = PHLTA,B327,22;%%
  %%CITATION = CMPHA,171,383;%%

\bibitem{Chamseddine}
  A.~H.~Chamseddine,
  ``Sl(2,C) gravity with complex vierbein and its noncommutative extension,''
  Phys.\ Rev.\  D {\bf 69}, 024015 (2004)
  [arXiv:hep-th/0309166].
  %%CITATION = PHRVA,D69,024015;%%

\bibitem{Wessgroup}
P.~Aschieri, C.~Blohmann, M.~Dimitrijevi\' c, F.~Meyer, P.~Schupp
and J.~Wess, {\it A Gravity Theory on Noncommutative Spaces},
Class.\ Quant.\ Grav. {\bf 22}, 3511-3522 (2005),
[hep-th/0504183];
P.~Aschieri, M. Dimitrijevi\' c, F.~Meyer and  J.~Wess, {\it
Noncommutative Geometry and Gravity}, Class.\ Quant.\ Grav. {\bf
23}, 1883-1912 (2006), [hep-th/0510059].

\bibitem{Estradaetal}
  S.~Estrada-Jimenez, H.~Garcia-Compean, O.~Obregon and C.~Ramirez,
  ``Twisted Covariant Noncommutative Self-dual Gravity,''
  Phys.\ Rev.\  D {\bf 78}, 124008 (2008)
  [arXiv:0808.0211 [hep-th]].
  %%CITATION = PHRVA,D78,124008;%%

\bibitem{MilanogroupG2}
  S.~Cacciatori, A.~H.~Chamseddine, D.~Klemm, L.~Martucci, W.~A.~Sabra and D.~Zanon,
  ``Noncommutative gravity in two dimensions,''
  Class.\ Quant.\ Grav.\  {\bf 19}, 4029 (2002)
  [arXiv:hep-th/0203038].
  %%CITATION = CQGRD,19,4029;%%

\bibitem{Banados}
  M.~Banados, O.~Chandia, N.~E.~Grandi, F.~A.~Schaposnik and G.~A.~Silva,
  ``Three-dimensional noncommutative gravity,''
  Phys.\ Rev.\  D {\bf 64}, 084012 (2001)
  [arXiv:hep-th/0104264].
  %%CITATION = PHRVA,D64,084012;%%

\bibitem{MilanogroupG3}
  S.~Cacciatori, D.~Klemm, L.~Martucci and D.~Zanon,
  ``Noncommutative Einstein-AdS gravity in three dimensions,''
  Phys.\ Lett.\  B {\bf 536}, 101 (2002)
  [arXiv:hep-th/0201103].
  %%CITATION = PHLTA,B536,101;%%

 \bibitem{AC1}
  P. Aschieri, L. Castellani,
  ``Noncommutative D=4 gravity coupled to fermions," DISTA-UPO/09,
   [arXiv:0902.3817 [hep-th]].

\bibitem{MilanogroupSG3}
  S.~Cacciatori and L.~Martucci,
  ``Noncommutative AdS supergravity in three dimensions,''
  Phys.\ Lett.\  B {\bf 542}, 268 (2002)
  [arXiv:hep-th/0204152].
  %%CITATION = PHLTA,B542,268;%%

  \bibitem{SW}
N.~Seiberg and E.~Witten,
{\it String theory and noncommutative geometry},
JHEP  {\bf 9909}, 032 (1999)
[hep-th/9908142].

\bibitem{Jurco}
  B.~Jurco, S.~Schraml, P.~Schupp and J.~Wess,
  ``Enveloping algebra valued gauge transformations for non-Abelian gauge
    groups on non-commutative spaces,''
  Eur.\ Phys.\ J.\  C {\bf 17}, 521 (2000)
  [arXiv:hep-th/0006246].
  %%CITATION = EPHJA,C17,521;%%

  \bibitem{Mielke}
  E.~W.~Mielke and A.~Macias,
  ``Chiral supergravity and anomalies,''
  Annalen Phys.\  {\bf 8}, 301 (1999)
  [arXiv:gr-qc/9902077].
  %%CITATION = ANPYA,8,301;%%

  \bibitem{ACD}
  P.~Aschieri, L.~Castellani and M.~Dimitrijevic,
  ``Dynamical noncommutativity and Noether theorem in twisted $\phi^{\star 4}$ theory,''
  Lett.\ Math.\ Phys.\  {\bf 85}, 39 (2008)
  [arXiv:0803.4325 [hep-th]].
  %%CITATION = LMPHD,85,39;%%

  \bibitem{book}
 P.~Aschieri, M. Dimitrijevic, P. Kulish, F. Lizzi, J. Wess,
 ``Noncommutative Spacetimes", Lecture Notes in Physics, vol. 774, Springer 2009.

   \bibitem{Aschieri}
P.~Aschieri,
``Noncommutative symmetries and gravity,''
J.\ Phys.\ Conf.\ Ser.\ {\bf 53} (2006) 799
[arXiv:hep-th/0608172].
%%CITATION = 00462,53,799;%%



\end{thebibliography}
\end{document}